\documentclass[preprint,tightenlines,eqsecnum,aps,epsfig]{revtex4}
\usepackage{epsfig}

\def\beq{\begin{equation}}   \def\eeq{
\end{equation}}

\begin{document}
\title{Tachyon dynamics of the black disc limit in the
hard small $x$ processes in  QCD.}
\author{ B. Blok\email{E-mail: blok@physics.technion.ac.il} }
\affiliation{Department of Physics, Technion---Israel Institute of
Technology, 32000 Haifa, Israel}
\author{ L. Frankfurt\email{E-mail: frankfur@lev.tau.ac.il} }
\affiliation{School of Physics and Astronomy, Raymond and Beverly Sackler
Faculty of Exact Sciences,
Tel Aviv University, 69978 Tel Aviv,
Israel}

\thispagestyle{empty}

\begin{abstract}
We investigate the effective field theory (EFT) which gives the
approximate description of the scattering of the two hard small
dipoles in the small $x$ processes in QCD near the black disc
limit(BDL). We argue that the perturbative QCD approaches predict
the existence of tachyon and visualize it in the approximation
where $\alpha'_P=0$ . We demonstrate that the high energy behavior
of the cross-section depends strongly on the diffusion law in the
impact parameter  plane. On the other hand, almost threshold
behavior of the cross section of the hard processes and
multiplicities, i.e. fast increase of cross sections (color
inflation),  melting of ladders into color network and softening
of the longitudinal distributions of hadrons  are qualitatively
insensitive to the value of  diffusion in the impact parameter
space. We evaluate $\alpha'_P$ near the black disk limit and find
significant $\alpha'_P$ as the consequence of the probability
conservation.

 \end{abstract}
\maketitle

\setcounter{page}{1}
\section{Introduction}
\par
One of the distinctive features of the hadron collisions at high
energies is the increase of the cross sections with the energy
which is often parameterized as the intercept of the Pomeron
$\alpha_P(t=0) > 1$, cf. review in ref. \cite{S.Donnachie and
P.Landshoff}. It follows from the unitarity of the $S$ matrix and
the analytic properties of amplitudes in the  momentum transfer
plane that the cross sections of hadron-hadron collisions can not
exceed the limit $\sigma\le c_h\ln^2(s/s_o)$
\cite{Heisenberg,Froissart}, with the coefficient $c_h$ determined
by the radius of the pion cloud of a hadron \cite{AMarten}. At
achievable energies the contribution of the Pomeron exchange
instead of the projectile scattering of the pion cloud dominates
in the peripheral collisions because pion is a pseudo-goldstone
meson of spontaneously broken chiral symmetry \cite{Zhalov}, cf.
analysis of the role of pion cloud in high energy processes in 
ref. \cite{Kharzeev}. The account of this
property of QCD explains why the observed value of $c_h$ is
significantly smaller than that arising from the black scattering
off the pion periphery of a hadron \cite{Zhalov}. Moreover, in the
case of the interaction increasing with energy the complete
absorption for the central hadron-hadron collisions at ultra-high
energies arises as the consequence of the numerically large and
increasing with the energy number of constituents in the wave
functions of the colliding hadrons. At ultrahigh energies the
coefficient $c_h$ should  be the same for all hadrons and nuclei
\cite{Zhalov}.

The  modelling of the hadron collisions at high energies within
the eikonal approximation where single Pomeron exchange is used as
"potential" indicated that the unitarity limit  should be achieved
at collider energies since the intercept of the Pomeron is
$\alpha_P(t=0)\> 1$, \cite{TerMartirosyan,Kaidalov,MBlok}. The
analysis of data on elastic pp collisions found that the
scattering at  central impact parameters achieves unitarity limit
at energies of FNAL \cite{Islam,MBlok}. At the domain of energies
to be achieved at LHC blackness of pp interaction will occupy
significantly wider region in the impact parameter space.

It seems now that the discussed above pattern of behavior of
amplitudes of high energy processes is similar to that expected at
sufficiently small $x$ in perturbative QCD for the cross sections
of hard processes. One of the striking properties of the hard
processes in the perturbative QCD is the rapid increase of
interaction with energy
\cite{Gross-Wilczek,Dokshitser,BFKL,Ciafaloni}. Such behavior can
not continue for arbitrary small $x$, otherwise it will violate
the probability conservation.  It was understood recently  that
this property of perturbative QCD and increase with energy of the
number of constituents within a projectile lead to complete
absorption for the central colorless dipole -hadron collisions.
cf.\cite{FSW}. Thus at ultra-high energies the total cross sections of
deep-inelastic scattering (DIS) should achieve the unitarity limit
\cite{Martin} : $\sigma({\rm dipole +Nucleon})\rightarrow c_D
\log^3(x_0/x)$ with the universal coefficient at ultra-high
energies. (Additional $\log x_0/x$ as compared to the hadronic
cross sections is due to the ultraviolet divergence of the
electromagnetic charge, cf. also ref.\cite{Gribov}). Both
theoretical calculations, of the elastic scattering of colorless
gluon dipole  and the analysis of the data on hard diffraction in
DIS observed at HERA, indicate that the black disc regime (BDL)
seems to be achieved for gluon distributions (but not for quark
distributions) at $Q^2~\sim {\rm few GeV}^2$ within the proton at
the verge of the kinematics of HERA. The blackness of interaction
including the hard one will be important in the central pp
collisions in the significantly wider region  in  the kinematics
of LHC, where it will compete with the hadronic final states in
the new particles production, for the review and references see
ref. \cite{FSW}.

In the previous paper \cite{BF} we deduced within the WKB
approximation Lagrangian of effective field theory (EFT)
describing the interactions of quasiparticles-pQCD color neutral
ladders relevant for small $x$ behavior of single scale hard
processes in the vicinity of BDL. This Lagrangian accounts for the
perturbative QCD calculations of small $x$ behavior of amplitudes
as an input. We found  that the perturbative QCD dynamics of the
single scale hard processes at small $x$ predicts the presence of
tachyon type behavior of amplitudes of  physical processes in EFT.
In particular, the increase with distance  of the matrix elements
of the correlators between currents evaluated within the
perturbative QCD indicates instability of produced pQCD system
which is an  important feature of the onset of the critical
phenomena  \cite{BF1}. (Note however that part of the increase of
correlator  with distance  follows from the Lorentz slow down of
interaction . This phenomenon resembles turbulence \cite{BF1}).
The existence of the tachyon leads to the spontaneously broken
continuous symmetries and related variety of critical phenomena.

At the intermediate stage of our analysis we were able to  use
similarity between the EFT and the model for high energy hadron
collisions based on preQCD Reggeon Calculus with the intercept
$\alpha_P(t=0) >1$ thorougly analyzed in cf. refs.
\cite{amati1,amati2,amati3,amati4,amati5}. Reformulating to hard 
processes in QCD near the black disk regime the results, obtained
within this model, we found kinks, spontaneous violation of
translational invariance and related two dimensional "phonons". At
the same time the finding that within the EFT the pQCD ladders are
overlapped in space and time rather often necessitates to go
beyond the approximations leading to EFT and to account for the
the interchange by constituents between overlapping ladders .
Account of exchange by color leads to color network - effect
beyond EFT. Thus a variety of new physical phenomena which are
absent in refs. \cite{amati1,amati2,amati3,amati4,amati5} have
been found in ref. \cite{BF}: melting of pQCD ladders and
formation of color network, importance of tunneling transitions
$\propto exp (-1/\alpha_s)$ .

In the approach developed in ref. \cite{BF} the important role is
played by the "Pomeron" slope $\alpha'_P$ that has not been
calculated in pQCD so far. Indeed, the  slow convergence of the
pQCD series at small x, cf. refs. \cite{Lipatov-Fadin,Ciafaloni},
precludes the reliable pQCD calculation of the value of
$\alpha'_P$. Besides, the significant value of $\alpha'_P$ arises
near BDL because of the fast increase with energy of pQCD ladder.
For the estimates of $\alpha'_P$ near the black disk regime  see
discussion below and ref. \cite{FSW}.

To visualize the similarities between small $x$ physics in QCD and
critical phenomena and the importance of V.Gribov diffusion in
small $x$ regime we consider in this paper the alternative
approximation $\alpha_P'=0$ and compare the obtained results with
those obtained in ref. \cite{BF}. The evident advantage of the
approximation $\alpha'_P=0$ is the possibility of the analytical
solution that made it possible to visualize the origin of the new
QCD phenomena found in ref. \cite{BF}. Note that because of a
dynamical generation of $\alpha'_P\ne 0$ near the BDL, in the
physical situation one can not put $\alpha_P'=0$ in eq. \ref{lag}.
However the analysis that equates $\alpha'_P$ to zero will give a
useful information on the dependence of different properties of
the theory on the theoretical input.

A number of alternative approaches to the ultra small x behavior
of hard processes  has been suggested and developed in the
literature, first of all the  color condensate approach
\cite{MV,JKLW,Mueller,McLerran,McLerran1} and related unitarization 
of the perturbative QCD contributions within the the
eikonal approximations,  see i.e. refs.
\cite{Kovchegov-Balitsky,Maor}. In difference with these
approaches we accounted for in ref. \cite{BF} ladder loops, found
physical vacuum within the WKB approximation and accounted for the
tunneling transitions between the physical and the perturbative
QCD vacua.  This lead to the  existence of the variety of new QCD
phenomena cf. ref. \cite{BF} such as melting of ladders and
formation of color network, spontaneous violation of two
dimensional translation symmetry and existence of two dimensional
"phonons", color inflation. Note that the black disk limit
behavior  does not follow from the color glass condensate
approach, at least if multiladder loops are neglected
\cite{Kovner-Wiedemann}. In addition, let us note that the color
glass condensate approach as well as the other leading order BFKL related
approaqches, are characterized by diffusion to small distances in
the transverse plane at high energies, i.e. the color condensate
is contained in narrow threads. On the other hand, in EFT
the diffusion to large distances dominates, and the high energy
densities may arise due to the rapid increase of the number of
constituents.
\par
Technically the effective field theory of ref. \cite{BF} has been
derived for one-scale processes such as the scattering of two
equally small size dipoles, where the evolution of scale is a
correction that we will neglect.
\par
This theory is governed by the  Hamiltonian \cite{BF}
\begin{eqnarray}
L&=&1/2(p\partial_y q -q\partial_y p) -\alpha'p\triangle_b q-
\mu pq-\kappa pq(p+q)\nonumber\\[10pt]
&-& c_{\rm dipole}\int \exp(-bQ/2) q(y,\vec B-\vec b)d^2 b \delta(y+Y)\nonumber\\[10pt]
&-&c_{\rm dipole}\int \exp(-bQ/2) p(y,\vec B-\vec b)d^2b \delta(y -Y),
\nonumber \\[10pt]
\label{lag}
\end{eqnarray}
Here $p=i\psi^+(y,\vec b),q=i\psi (y,\vec b)$ are operators of
production and annihilation of the pQCD ladders , $y=\log(x_o/x)$
is rapidity, $\vec b$ is the impact parameter, $\mu$ is a
"Pomeron" intercept, and $\kappa$ is a three-ladder vertex.  We
took $\kappa$ to be constant near the BDL which properly accounts
for hard QCD physics. On the contrary within the LO BFKL
approximation $\kappa$ is singular as a function of the momentum
transfer $t$  \cite{Bartels-Ryskin}. Such a bizarre behavior
demonstrates that physics of large transverse distances
artificially suppressed within LO BFKL approximation gives
actually significant contribution within this approximation. This
is because within LO BFKL approximation there is no separation of
scales between soft and hard QCD phenomena , no running coupling
constant and no Sudakov form factors.  The value $c_{\rm dipole}$
is determined by the normalisation of the dipole wave  function.

In ref. \cite{Lipatov} the slope $\alpha'_P$ has been put to zero
from the beginning. The latest NLO calculations \cite{Ciafaloni}
show that the rapid diffusion to large nonperturbative distances
characteristic for the leading order BFKL approximation is
suppressed when the next to leading order approximation is
accounted for. However the  randomness of the gluon radiation and
the related Gribov diffusion in impact parameter space should be
there \cite{Gribovdiffusion}.  We will find significant value for
$\alpha'_P$  in the kinematics near the black disk limit, see
discussion below.

\par
To visualize the sensitivity of QCD phenomena to the value of
$\alpha'_P$ it is interesting to investigate what will happen if
to  put it to zero . This will show quantatively, how the
asymptotic behavior of the cross-section is influenced by the
diffusion in the transverse plane. As we shall see shortly, the
diffusion in a transverse plane plays a crucial role. For example,
the account of V.Gribov diffusion in the impact parameter space
leads to Froissart behavior \cite{BF}, while as we shall see below
the absence of diffusion leads within EFT to grey disc-asymptotic
constant value of the cross-sections. Physical cross section still
increases  with energy because of effects neglected in EFT.
 \par
The paper is organized in the following way. In section 2 we
formally consider the case $\alpha'_P=0$ to visualize presence of
tachyon, to study the qualitative effect of V.Gribov diffusion in
impact parameter space on the high energy behavior of structure
functions. In section 3 we argue that physical value of
$\alpha'_P$ is significant and can not be  neglected at high
energies. We summarize our results in the conclusion.

\section{Effective Field Theory for small size dipole scattering
without diffusion.}

\par
The Lagrangian of the EFT without V.Gribov diffusion in impact parameter space
is given by eq. \ref{lag}, with $\alpha_P'=0$. Thus the Lagrangian is
\begin{eqnarray}
L&=&1/2(p\partial_y q -q\partial_y p) -
\mu pq-\kappa pq(p+q)\nonumber\\[10pt]
&-& c_{\rm dipole}\delta (y) qf\nonumber\\[10pt]
&-&c_{\rm dipole}p \delta(y -Y)\nonumber\\[10pt]
\label{lag2}
\end{eqnarray}
Recall that the first term accounts for the kinetic energy, the second
corresponds to "Pomeron" intercept $1+\mu$. The vertex $\kappa$ is
effectively a 3-reggeon vertex.

\par
One approach to the theory is to solve the  classical  equations of motion that
follow from the Lagrangian \ref{lag2} and then consider the quasiclassical
fluctuations around this solution.
\par
The equations of motion for the Hamiltonian corresponding to eq. \ref{lag2} can be
easily solved. Indeed, these equations of motion can be written as
\beq
\frac{dp}{dy}=-\frac{dH}{dq}\,\,\,\,\,\frac{dq}{dy}=\frac{dH}{dp}
\label{ham}
\eeq
where
\beq
H=-\mu pq+\kappa pq(p+q)
\label{hami}
\eeq
Since the energy is a constant of motion, we can put H=E, obtain p=p(q,E), and then
explicitly solve the second of equations \ref{ham} for q. There are  four critical
points for this action: $$(p,q)=(0,0), (0,\mu/\kappa), (\kappa/\mu ,0)$$ with $E=0$, and
$$(p,q)=(-\mu/(3\kappa),-\mu/(3\kappa))$$ with $E<0$. The S-matrix
in the classical approximation is given by
\beq
S\sim exp(-A_{\rm cl})
\label{S1} \eeq
where we need to calculate the classical action $A_{\rm cl}$ for each value of E,
then take the minimal one, and consider fluctuations around it. It turns out that the
dominant contribution arises from the trajectories with
$E\rightarrow 0$, that corresponds to the trajectories connecting
the perturbative and nonperturbative vacua with E=0. (we refer the
reader to the ref.\cite{amati6} for the detailed analysis of the
trajectories). These trajectories are the quantum mechanical
analogues of the kinks of the 2+1 dimensional theory \cite{BF}.
Evidently this approach can be easily extended to the arbitrary number of
dimensions cf. cite{BF}.

\par
The approximation  \ref{lag2} is a nonhermitean quantum mechanics.
It can be reformulated by substituting variables into Hermitian
quantum mechanics. After the transformation: $q=x^2, p=d/d(x^2)$ ,
and transformation of the Hamiltonian \beq
H=\exp(-F(x)H_0\exp(F(x), \label{p} \eeq where \beq
F(x)=(1/4)(x^2-\mu/\kappa^2)^2, \label{f} \eeq we obtain the
Hermitean Hamiltonian cf. \cite{amati5,amati7} \beq
H=-\kappa/4\frac{d^2}{dx^2}+V(x) \label{H} \eeq where \beq
V(x)=(\kappa/4)(x^2(x^2-\mu/\kappa)^2+\mu/\kappa-x^2) \label{V}
\eeq This Hamiltonian determines the same Borel convergent series
for $\mu < 0$ as the original one . So both theories are
equivalent, at least for the low energy levels we are interested
for. The derived interaction in the Hermitian quantum mechanics is
the double-well (Higgs-type) potential .  In the case $\mu>0$
predicted by the perturbative QCD the theory undergoes the
transition similar to a Higgs phenomena: there are two degenerate
vacua (see Fig 1), while the perturbative vacuum corresponds to
the maximum of V and a tachyon. Double well structure of the
interaction clearly visualizes the origin of the tachyon predicted
by the pQCD calculations, the similarity between the BDL in hard
QCD and second order phase transitions. The spontaneous violation
of the continuous symmetries, in the case when $\alpha'_P$ is kept
nonzero, makes the analogy with second order phase transitions
even more close.

\par
The S-matrix is determined by
\beq
S(f,Y)=<0\vert \exp(-c_{\rm dipole}p)\exp(-H_0Y)\exp(-c_{\rm dipole}q)\vert 0>
\label{s}
\eeq
in terms of the original nonhermitian hamiltonian and can be expressed in
terms of the eigenfunctions and eigenvalues of the equivalent hermitean
hamiltonian \cite{amati5,amati7}:
\beq
S=\frac{\sum_n<\phi_0\vert \exp(-F)O_2 \exp(F)\vert\phi_n><\phi_n\vert
 O_1\vert\phi_0>\exp-(E_n -E_0)Y}{<\phi_0\vert
 \exp(F)O_2\exp(F)\vert\phi_n><\phi_n\vert O_1\vert\phi_0>},
\label{S2} \eeq where F(x) is given by the equation \ref{f},
$$O_1=\exp(-cx^2),$$
$$O_2=\exp(+c\frac{\delta}{\delta_{x^2}}),$$
 and $<\phi_0\vert
O\vert \phi_n>$ means $\int^{\infty}_{-\infty}\phi_0(x)O\phi_n(x)
dx$.

\par
We shall briefly formulate here the main results which  are the same for both approaches
discussed above:

\par A)
The cross section of hard processes in high energy limit  is constant. The value
of the constant depends on the relation between the source strength and
the ratio $\mu/\kappa$.
\par
We consider in the paper weak coupling limit for single scale hard
processes $\alpha_s N_c<<1$ where, the cross-section tends to
\beq
\sigma_{Y\rightarrow \infty}\exp(-c_{\rm dipole}\mu/\kappa)\sim
{\rm const}
\label{lag4}
\eeq
In other words, one gets instead of Froissart limit ( $\sigma (s)\sim \log^2(s)$
behavior) a grey disk behavior.

\par B)
Moreover,  the tunnel transitions in this case lead (albeight at
extremely high $y \sim \exp(\mu^2/\kappa^2)$ energies) to
slow decrease of the cross-section. Indeed, the S-matrix  is controlled
by the lowest energy level in this double-well. This level is aroused due to
nonperturbative effects
and is $ E_1\sim exp(-\mu^2/(2\kappa^2))$, thus the cross-section has a plato
for $$\exp(\mu^2/(2\kappa^2))>>y>>1/\mu .$$.  The cross-section in this
plato is given by eq. \ref{lag4}. For larger rapidities
\beq
 \sigma\sim 1/s^{\exp(-\mu^2/(2\kappa^2))}
\label{lag5} \eeq The approach to the grey disk limit corresponds
to "color inflation" phenomena, argued in ref.\cite{BF}. Indeed
\cite{amati5} the cross-section is well approximated by: \beq
\sigma \sim c_{\rm dipole}^2\exp(\mu Y)/(1+\exp(c_{\rm
dipole}\mu/\kappa))\exp(\mu Y) \label{lag6} \eeq In other words
the cross-section shows a step like behaviour with a step width of
order $1/\mu$ in rapidity space (Fig. 2).
\par
Note that grey disc limit has academic interest only since in
realistic case $\alpha'_P$ can not be neglected near BDL  because
of renormalization due to tachyon effects,see the discussion
below.

\par
In the approximation where $\alpha'_P=0$ all ladders are located at the same impact
parameters. So overlap of ladders found in \cite{BF} should be even more valid in this
case. Thus melting of overlapping ladders due to exchange by constituents and
formation of  color network are
important even in the case when diffusion in impact parameter space is neglected.

\section{The evaluation of $\alpha'_P$ near the black disk limit. }

In the kinematics near the black disk limit where the partial
waves become close to 1 the large value of $\alpha'_P$ becomes
important because of the rapid increase with energy of pQCD
amplitude. This restriction due to the probability conservation
has not been considered in the formulation of  EFT in ref.
\cite{BF} where $\mu,\alpha'$ were parameters of a bare "Pomeron".
The requirement of the self-consistency of the approximation put
strong restriction on the value of the renormalized $\alpha'_P$.
The general expression for partial wave is $Im f=cs^{\mu} exp
(-bQ)$. Here $b$ is the impact parameter, $Q$ is the scale of the
discussed processes. For the sake of estimate we assumed that for
the peripheral scattering the pQCD ladder should dominate. Since
${\rm Im} f \le1$ the important impact parameters can be evaluated
from the condition ${\rm Im} f=1$ which is achieved within the
EFT-cf. \cite{BF}.
 At $x=Q^2/s\to 0$  this gives: $b_{\rm max}Q=\ln(s/Q^2)$.
Here $b_{\rm max}$ is the maximal impact parameter b for which $Im f=1$.
Thus $\alpha'_P$ defined as the derivative of the slope of t dependence of 
the square of the    amplitude is given by the formulae:
 $\alpha'_P\equiv 1/2 (d/d\ln(s/s_o)) b_{max}^2 $ cf. \cite{FSW}.
Thus near the black disk limit the  renormalized $\alpha'_P$ for
hard processes should be even larger than that  for soft QCD
processes. So the limit $\alpha'_P=0$ is unphysical , but it helps
to visualize different aspects of QCD physics relevant for the
black disk regime.

\section{Conclusion.}
\par
We demonstrate that the asymptotic behavior of the high energy
processes strongly depends on the diffusion in the impact
parameter space. On the other hand, the sharp increase of the
multiplicities and of the cross section the during transition from
the perturbative to the asymptotic regime, melting of pQCD ladders
and formation of color network, and  the softening of the forward
distributions over the longitudinal momenta seem to be  a
diffusion-independent phenomena.
\par
We argue that in a physical case of the hard energy scattering in
QCD near the black disk limit the $\alpha'_P$ value can not be
small , because of the renormalization of the parameters of the
effective Lagrangian due to the tachyon effects (the phenomenon
neglected in refs. \cite{amati1,amati2,amati3,amati4,amati5,
amati6,amati7,BF}. In other words the diffusion in impact
parameter space can never be neglected for physical problems,
although it is legitimate to investigate it's influence on the
cross-sections in the approximate approaches .

\newpage
\begin{figure}[htbp]
\centerline{\epsfig{figure=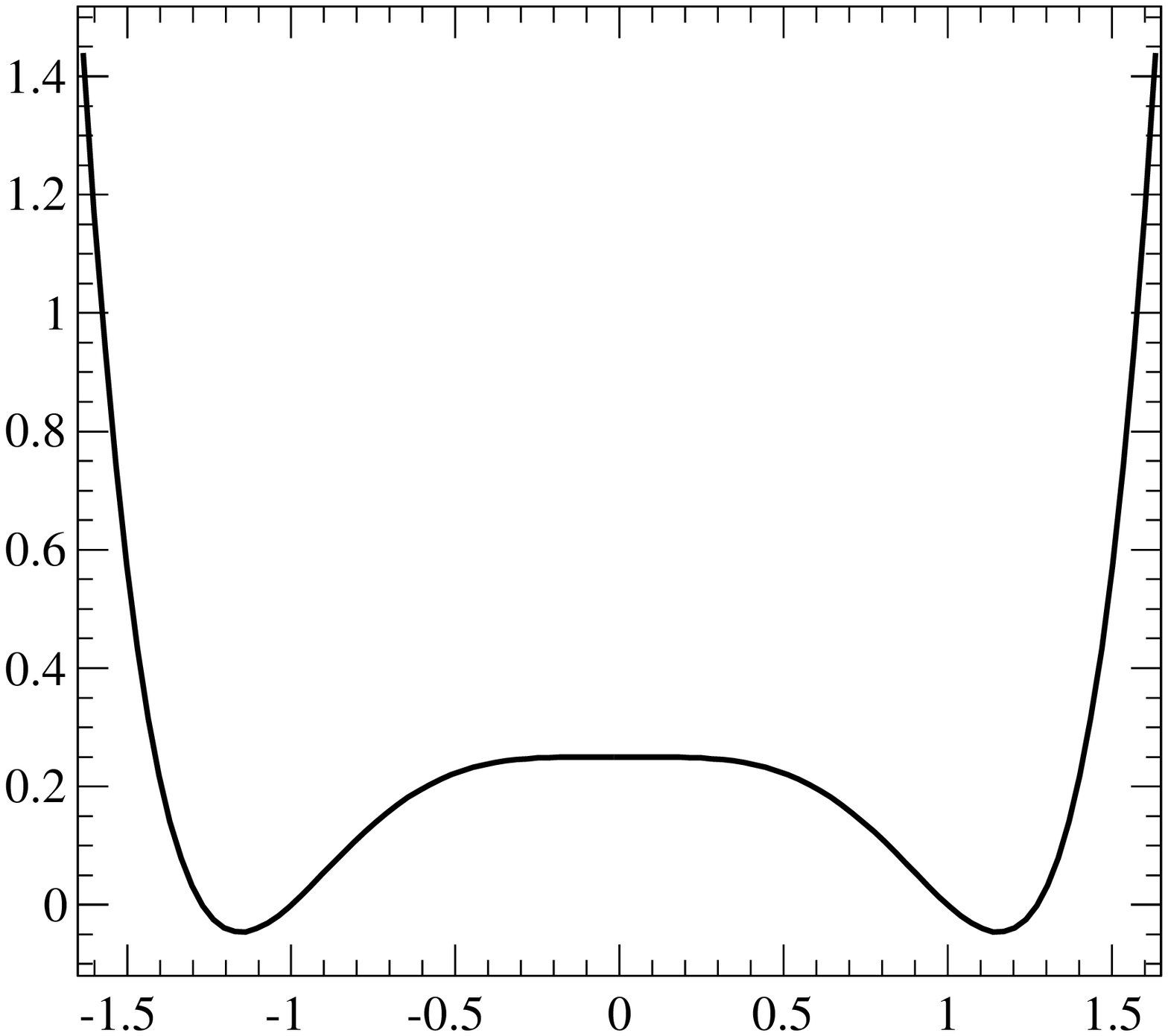,width=15cm,height=15cm,clip=}}
\caption{The effective potential for $\mu>0$ (tachyon is
present)}
\label{SK2}
\end{figure}
\begin{figure}[htbp]
\centerline{\epsfig{figure=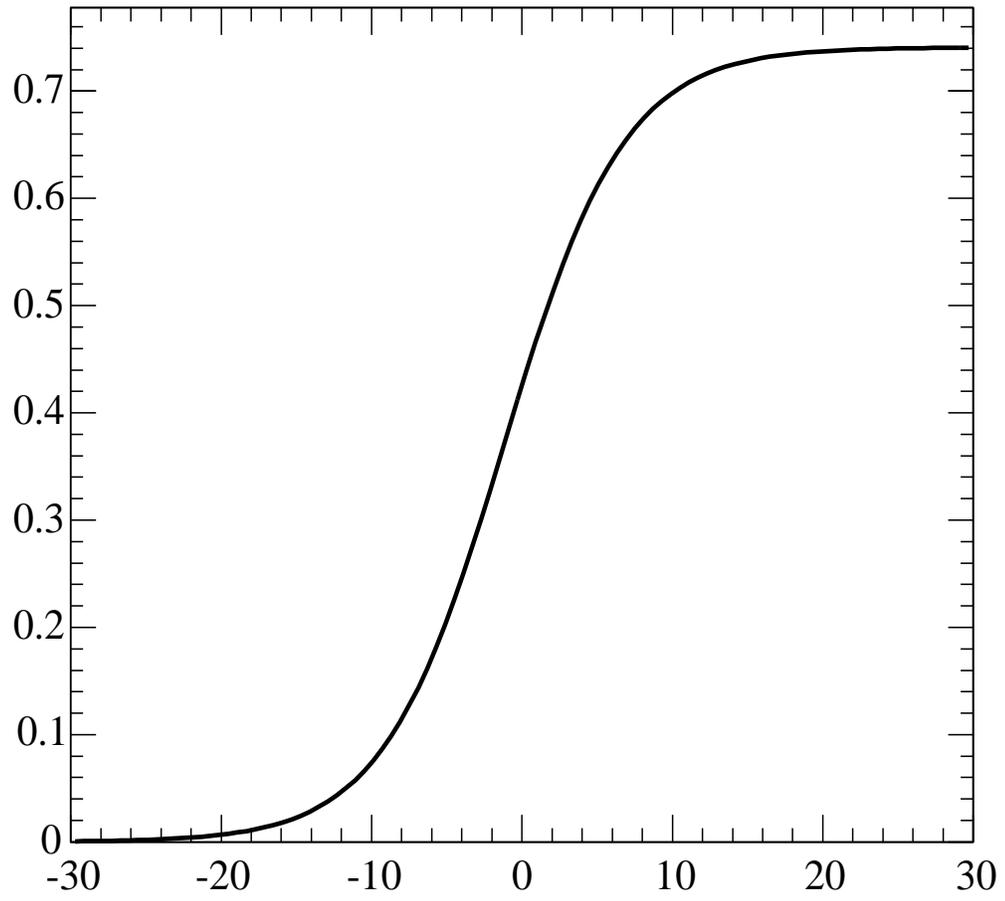,width=15cm,height=15cm,clip=}}
\caption{ The cross-section dependence on rapidity Y (color
inflation)} \label{SK3}
\end{figure}
\end{document}